# On the Path to 6G: Embracing the Next Wave of Low Earth Orbit Satellite Access

Xingqin Lin[†], Stefan Cioni[‡], Gilles Charbit[§], Nicolas Chuberre[⊥], Sven Hellsten[†], and Jean-Francois Boutillon[⊥]
[†]Ericsson, [‡]European Space Agency, [§]MediaTek, [⊥]Thales Alenia Space
Contact: xingqin.lin@ericsson.com

*Abstract*— Offering space-based Internet services with mega-constellations of low Earth orbit (LEO) satellites is a promising solution to connecting the unconnected. It can complement the coverage of terrestrial networks to help bridge the digital divide. However, there are challenges from operational obstacles to technical hurdles facing the development of LEO satellite access. This article provides an overview of state of the art in LEO satellite access, including the evolution of LEO satellite constellations and capabilities, critical technical challenges and solutions, standardization aspects from 5G evolution to 6G, and business considerations. We also identify several areas for future exploration to realize a tight integration of LEO satellite access with terrestrial networks in 6G.

## I. THE NEW SPACE RENAISSANCE

As the commercial fifth generation (5G) systems are being rolled out throughout the world, research focus starts to be gradually shifted towards the sixth generation (6G) systems that are anticipated to achieve another giant leap in communications [1]. We expect that the communication with mega-constellations of low Earth orbit (LEO) satellites will be one of the connectivity's new frontiers on the path to 6G, complementing terrestrial networks to provide limitless connectivity everywhere, thus helping to bridge the digital divide [2].

An LEO is an Earth-centered orbit with an altitude between 350 km and 2,000 km above sea level. Compared to medium Earth orbit (MEO) or geosynchronous Earth orbit (GEO) counterparts, LEO's proximity to Earth results in lower latency in LEO satellite access, less energy for launching, and less power for signal transmission from and to the satellites. However, LEO satellites at 600 km altitude travel at a speed of around 7.8 km/s. The fast movement of LEO satellites and the relatively limited coverage area of an individual LEO satellite require a large constellation of satellites to provide service continuity across the targeted coverage area [3]. Figure 1 illustrates the global coverage offered by an LEO constellation with hundreds of satellites at 600 km altitude.

Over the last several years, the world has witnessed a resurgent interest in space-based Internet services, particularly with mega-constellations of LEO satellites such as SpaceX Starlink, Amazon Kuiper, and OneWeb [4]. These initiatives are ambitious, involving thousands and tens of thousands of satellites and driven by non-traditional, cash-rich, and high-tech players. The interest in LEO satellite access has run its course before. A series of proposals on providing connectivity from space surged in the 1990s, but most of them failed, and a few (e.g., Iridium [5] and Globalstar [6]) have had varying degrees of success.

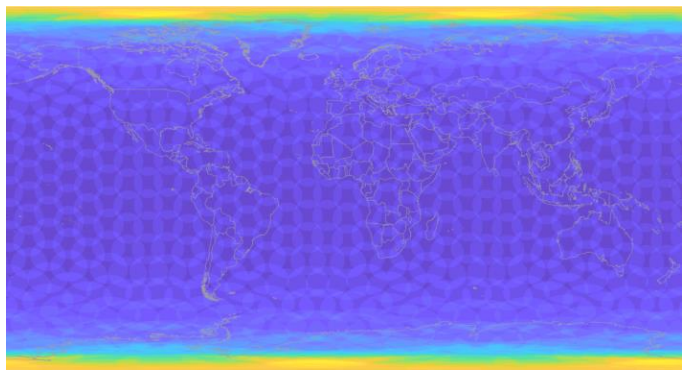

**Figure 1**: Global coverage of a LEO constellation with hundreds of satellites at 600 km altitude. (The colors indicate the number of satellites in view for each point on Earth. Dark blue denotes only one satellite in view. The lighter the color, the higher the number of satellites in view).

Several factors have been contributing to the renewed interest in LEO satellite access. A key enabler is a significant decrease in launch cost with the advent of disruptive launchers reusing rocket parts, as provided, for example, by SpaceX. Meanwhile, the use of components off the shelves (COTS) and the adoptions of lean principles in satellite design and manufacturing allow for mass production with a faster manufacturing cycle at reduced costs. It has also become commercially feasible to use advanced technologies in satellite communications such as multi-spot beam technologies and sophisticated onboard digital processing. Another key driver is a greater willingness to invest in LEO satellite access to help connect the unconnected, motivated by commercial potential, economic development, and humanitarian considerations of bridging the digital divide.

The 3rd Generation Partnership Project (3GPP) has been working on adapting 5G systems to support satellite communications [7][8]. Evolving 5G to support LEO satellite access is built on the inherent flexibility of the 5G systems. The first design objective is to connect 5G handset devices by 3GPP-based satellite access networks in the sub-6 GHz spectrum, so that 5G connectivity can be provided to the areas where terrestrial 5G networks are not available. The second design objective is to provide broadband connectivity to more advanced devices, such as very small aperture terminal (VSAT) or Earth station in motion (ESIM), especially in higher frequencies (e.g., Ku/Ka bands). Besides direct satellite access, 3GPP has also been working on 5G satellite backhaul, which can facilitate the offering of cellular services in the areas where terrestrial backhaul means are not possible or prohibitively expensive to build.

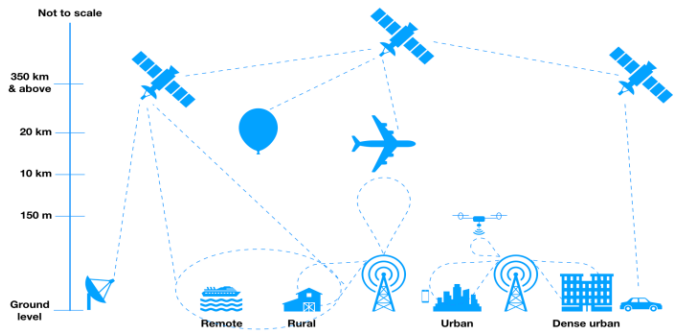

**Figure 2: Ubiquitous 6G coverage provided by tightly integrating terrestrial networks with LEO satellite access.**

5G was initially designed targeting terrestrial communications without considering LEO satellite access. We expect that there will be only moderate specification enhancements to enable 5G to support LEO satellite access, leading to suboptimal performance. In addition, most of the advanced 5G features will likely be out of reach for LEO satellite access in the 5G era. The integration of 5G terrestrial networks with LEO satellite access will be loose. In contrast, we anticipate that the integration will become much tighter in 6G, providing seamless mobility between terrestrial and LEO satellite access networks. As illustrated in Figure 2, the tight integration of terrestrial networks with LEO satellite access will be essential to achieve global coverage in the 6G era. Figure 2 also illustrates the overall LEO satellite access network architecture, which consists of feeder link connecting gateway and satellite, service link connecting satellite and terminal, and inter-satellite link (ISL) connecting one satellite to another.

The excitement about space-based Internet services is often clouded by myths and confusion. There are challenges, from technology to business model, facing the development of LEO satellite access with mega-constellations. The objective of this article is to provide an overview of state of the art in critical aspects of LEO satellite access to shed light on the research and development in this area on the path to 6G. In this article, we focus specifically on LEO constellations, but it shall be noted that GEO and MEO satellite systems (e.g., O3b) can also play an essential role in providing space-based Internet services with different technical constraints and design trade-offs.

## II. EVOLUTION OF LEO SATELLITE CONSTELLATIONS AND CAPABILITIES

### A. LEO satellite constellations

The exemplary early LEO systems include Iridium [5] and Globalstar [6], for which the first generations were launched in the 1990s. The Iridium constellation was composed of 77 satellites with an altitude of 780 km (and hence the name Iridium, being the element with the atomic number equal to 77). The Globalstar constellation consisted of 48 satellites at approximately 1400 km altitude. Both satellite systems were designed to serve specific handset terminals with the second generation (2G) mobile services such as phone calls and short data messaging with a data rate of up to 14.4 kbps. Interestingly, the onboard satellite capabilities were quite advanced for that period, particularly for Iridium, since each spacecraft was equipped with four ISLs supporting communication with adjacent satellites at a data rate of 10 Mbps and high-speed processors for routing voice and data connections.

For many reasons (both technically and commercially driven), the LEO systems launched in the 1990s were not as successful as expected because they only provided voice and data connectivity to a limited number of user equipments (UEs). But a wind of change is blowing thanks to the wide range of 5G use cases and new market opportunities. There has been a resurgence of interest in using mega-constellations of LEO satellites to provide connectivity from space [9]. Some new constellations such as OneWeb, Starlink, AST SpaceMobile, Amazon Kuiper, and TeleSat have planned to launch hundreds or thousands of space vehicles to provide global connectivity, complementing existing terrestrial network infrastructures. Some of the key information on these prominent LEO constellations is summarized in Table 1. We refer the interested readers to [10] for a more in-depth discussion on constellation design.

The design of these new mega-constellations is driven by the target use case and market segment. In line with the available spectrum allocated to mobile satellite services below 6 GHz, an LEO satellite system will mainly focus on the Internet of Things (IoT) devices and services with low-medium data rates directly to handheld terminals. As far as satellite spectrum in higher frequencies is concerned (e.g., Ku/Ka bands), the new LEO satellite constellations will target broadband connectivity to more advanced devices, such as VSAT or ESIM.

### B. LEO satellite capabilities

LEO satellite payload should be sized according to the average target traffic, but the available platform power is often a constraint. This calls for efficient power recharging capabilities when the traffic is low (e.g., switching off most operational functions over poles and oceans to maximize the recharging phase with solar panels). The adoption of ISLs can reduce the overall number of gateways on the ground, minimize the infrastructure and operational costs, and serve oceans beyond coastal areas.

The technological constraints and design parameters in the two distinct spectrum ranges, i.e., sub-6 GHz and higher frequencies in, e.g., Ku/Ka bands, are different. Designing a satellite payload that covers both spectrum ranges is considered impractical in the foreseeable future. Hereafter, we describe some more specific satellite payload design aspects as a function of the selected frequency range.

Aiming at minimizing the impact on UE for use cases below 6 GHz, relatively large antenna in space is the very first enabling technology to offer comparable uplink and downlink received powers for the target devices. On the other hand, even though VSAT or ESIM devices can provide higher transmission power and antenna gain in higher frequencies than conventional mass-market handhelds, the challenges in space are not reduced. In particular, to cope with a diversified and non-uniform traffic demand on the ground and addressing the moving platform mounted devices (e.g., vessels and airplanes) or rooftop mounted devices (e.g., household), the satellite payload must guarantee efficient and flexible usage of the space resources as well as lean manufacturing and integration costs.

| Constellation | # of satellites [launched / planned] | Altitude [km] | Frequency allocation[1] | Terminal type |
|---|---|---|---|---|
| Iridium (1st gen) | 66 / 77 | 780 | 1.6 GHz (TDD) | Specific handset |
| Globalstar | 48 / 48 | 1440 | DL: 2.4 GHz; UL: 1.6 GHz | Specific handset |
| AST SpaceMobile | 1 / 240 | 720 | < 2 GHz | Commercial handset |
| OneWeb | 212 / 650 | 1200 | DL: 12 GHz; UL: 14 GHz | VSAT, ESIM |
| Starlink[2] | 1625 (100 inactive) / 1584 | 550 | DL: 12 GHz; UL: 14 GHz | VSAT, ESIM |
| TeleSat | 0 / 298 | 1015 and 1325 | DL: 20 GHz; UL: 29 GHz | VSAT, ESIM |
| Amazon Kuiper | 0 / 3200 | 600 | DL: 20 GHz; UL: 29 GHz | VSAT, ESIM |

Note 1: The downlink (DL) / uplink (UL) frequencies are indicative (rather than the specific allocated carrier frequencies).
Note 2: Starlink claims more than 12,000 satellites in different frequency allocations and orbit altitudes. Here we report the data related to the first-generation constellation operating in Ku-band at 550 km altitude.

Table 1: General information on some existing and planned LEO constellations (data collected in May 2021).

The technology readiness level (TRL) for many of the technology enablers is high. Indeed, Starlink has more than 1600 LEO satellites in orbit already. The critical aspect of space-qualified technologies is the TRL for the specific design requirements. For example, onboard active antenna technologies are now available, as equipped in the Starlink satellite, which has individually shapeable and steerable beams. The number of beams in the Starlink satellite is configurable and can range from 8 to 32. But the next LEO mega-constellations might require larger arrays or higher numbers of antenna elements, which will call for further optimizations and integration in the payload.

Recently, a detailed analysis of 5G requirements towards the design of future satellite systems was provided in [10]. Some of the key enabling technological aspects are related to the provision of novel space-qualified radio frequency (RF) and microwave components, such as high-efficiency power amplifiers based on gallium-nitride (GaN) and silicon-germanium (SiGe) technologies to enable cost-efficient integration of RF structures. Other advanced techniques such as fully digital beamforming and beam hopping schemes are also essential factors to consider for the design optimization of the satellite payload as a function of the power, mass, and affordable cost budget for a specific target service and traffic demand.

Next-generation payload architectures will utilize active antennas and onboard digital processors to enable a highly dynamic flexibility for modifying the carrier frequencies, bandwidths per carrier, allocated powers per beam, and even changing the beam layouts almost instantaneously as well as jointly with beam-hopping techniques. This means that the radio resource management system needs to cope with many degrees of freedom in a highly constrained environment to find the best possible allocation of resources at a system level. Note that due to the dynamicity of the system with satellites moving relatively to the ground and each other, every allocation scheme almost needs to be recomputed each time. Appropriate radio resource management algorithms are key enablers for an LEO constellation to be efficient.

## III. KEY TECHNICAL CHALLENGES AND SOLUTIONS OF LEO SATELLITE ACCESS

The main challenges of LEO satellite access arise from the fast movement of LEO satellites, leading to time-varying propagation delays, large time-varying Doppler shifts, and specific cell patterns (quasi-Earth fixed or Earth moving cells). In this section, we discuss key solutions to these challenges.

To start with, we summarize example parameters for LEO satellites operating in the S band (i.e., 2 GHz) and at 600 km altitude in Table 2 based on the 3GPP TR 38.821 [11] and TR 36.673 [12]. We can see that the LEO satellite targeting handheld UEs (e.g., smartphones) has relatively high effective isotropic radiated power (EIRP) density and antenna gain-to-noise-temperature (G/T) figures. In contrast, an LEO satellite targeting IoT connectivity can be less powerful. We refer the interested readers to the 3GPP TR 38.821 [11] and TR 36.673 [12] for comprehensive link budget analysis and results in diverse scenarios with different types of satellites (which differ in transmit power, antenna gain, etc.), different types of devices, different frequency ranges, different altitudes, etc.

### A. Serving satellite cell access

The fast movement of an LEO satellite at 600 km altitude can result in a high Doppler shift in the order of 24 ppm (i.e., ±48 kHz at 2 GHz carrier frequency). The Doppler shift is higher than a typical initial UE oscillator inaccuracy which is in the order of 10 ppm. The frequency error that the UE observes in the downlink is a combination of the Doppler shift and frequency offset (due to crystal oscillator mismatch). A UE can acquire time and frequency synchronization in the downlink using synchronization signals despite the compounded frequency error. However, UEs in the same cell may tune their local frequency references to significantly different frequencies due to different Doppler shifts. If downlink signal frequency is used as a frequency reference for uplink transmission, uplink signals of different UEs would be frequency misaligned at the receiving network node. This frequency misalignment would lead to loss of uplink orthogonality in a radio access system utilizing orthogonal frequency division multiple access (OFDMA), which is the *de facto* multi-access scheme in modern wireless systems.

| Example satellite dimensioning | LEO satellite targeting IoT devices | LEO satellite targeting handheld UEs |
|---|---|---|
| Equivalent satellite antenna aperture diameter | 0.4 m | 2 m |
| Satellite EIRP density | 28.3 dBW/MHz | 34 dBW/MHz |
| Satellite Tx / Rx max. antenna gain | 16.2 dBi | 30 dBi |
| 3 dB beam width | 22.1 degree | 4.4 degree |
| G/T | -12.8 dB/K | 1.1 dB/K |
| Beam diameter at nadir[1] | 234 km | 46 km |
| Note 1: Beam diameter is calculated from 3 dB beam width assumption. | | |

**Table 2: Example parameters for LEO satellites operating in the S band and at 600 km altitude.**

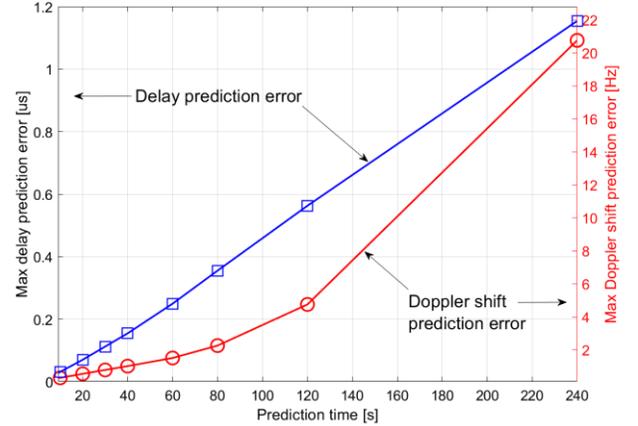

**Figure 3: Prediction accuracy of delay and Doppler shift versus prediction time for a LEO satellite operating in S band and at about 600 km altitude.**

To address the issue of frequency misaligned uplink transmissions, different frequency adjustment values at different UEs are needed in the uplink to compensate for their respective Doppler shifts due to satellite motion. Similarly, there is a need to align the uplink transmission timings by using different timing advance values at different UEs to compensate for their different propagation delays (ranging from a few milliseconds to tens of milliseconds for LEO satellite access) during initial random access procedure [13] as well as in connected mode when transmitting data traffic in the uplink (see, e.g., [11][14] for more details). Then, the central question is how to determine the appropriate frequency adjustment and timing advance at UE.

One solution, which is being standardized in 3GPP, is to equip the UE with a global navigation satellite system (GNSS) chipset. Thus, the UE can determine its position and velocity. The LEO satellite access network can broadcast satellite ephemeris-related data which conveys information about the serving satellite's position and velocity. A satellite orbit can be fully described using six parameters with different representations. With the known positions and velocities of the satellite and UE, the UE can determine the downlink and uplink Doppler shifts and the propagation delay of the service link. Based on the estimates, before initial access, the UE can adjust its uplink frequency and apply timing advance in uplink transmission, which helps to achieve multi-access orthogonality in the uplink. The movement of the serving LEO satellite requires the UE to continuously track the needed timing advance and frequency adjustment. For the LEO satellite targeting IoT connectivity in Table 2, the one-way delay drift and Doppler shift drift can be up to 46 $\mu s$/s and 641 Hz/s, respectively (which are larger than their counterparts in the LEO satellite targeting handheld UEs due to a larger beam size). Therefore, the UE needs to predict the satellite's position and velocity, propagation delay, and Doppler shift at a point in time different from the reference time associated with the satellite ephemeris.

The prediction based on satellite trajectory calculation, in general, degrades with the increasing age of the used ephemeris data for different reasons, including atmospheric drag, maneuvering of the satellite, imperfections in the used orbital models, etc. Figure 3 shows that the satellite's position and velocity can be predicted with good accuracy for deriving the needed timing advance and frequency adjustment in the order of tens of seconds or possibly minutes ahead in time. For example, for the prediction of 60 s ahead, the maximum radial satellite-UE delay error and Doppler shift error are 0.08 $\mu s$ and 4.8 Hz, respectively. This allows sufficient accuracy for adequate detection and demodulation performance within the cyclic prefix of orthogonal frequency division multiplexing (OFDM) waveform (e.g., 4.7 $\mu s$) and within the ±0.1 ppm of uplink transmission requirement (e.g., ±200 Hz at 2 GHz). We refer the interested readers to [14] for more results.

### B. Mobility management

Mobility management involves UEs in idle mode and UEs in connected mode. Idle mode mobility includes cell selection and reselection, tracking, and paging, while connected mode mobility mainly refers to handover. In terrestrial networks, cells are fixed, whereas UEs might be mobile with different trajectories. In contrast, cells move in LEO satellite access networks, leading to changes of UEs' serving cells in the order of several seconds (for Earth moving cells) to several minutes (for quasi-Earth fixed cells). This requires a rethinking of the mobility management procedures used in terrestrial networks.

To facilitate mobility management, UEs perform serving and neighboring cell measurements. The LEO satellite cells move over time in a predictable manner, as characterized by the satellite ephemeris data. The UE can utilize the ephemeris data to predict the trajectory of the LEO satellites over time and perform measurements at appropriate time instants. This solution is being standardized in 3GPP. With the measurement results, the UE can perform cell selection and reselection in idle mode and report the results to the network to assist with handover management in connected mode.

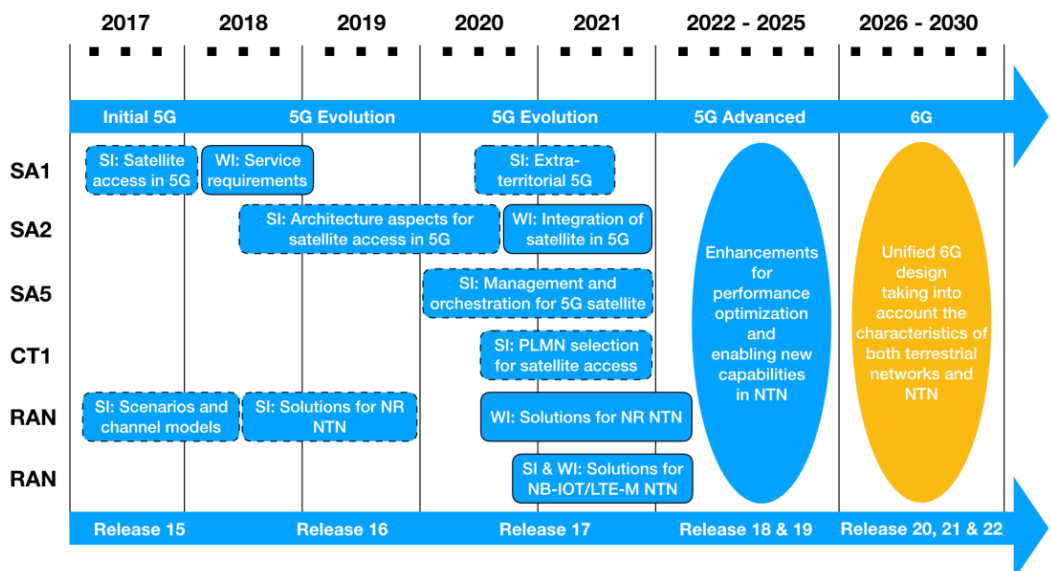

Figure 4: Satellite related activities in 3GPP roadmap (indicative).

The network needs to track a UE in the idle mode so that the network can page the UE in a prompt manner. To this end, the network provides the UE with a list of tracking areas. If the UE moves to a place that does not belong to any tracking area in the provided list, it needs to notify the network via tracking area update. In the LEO satellite access network with Earth moving cells, if the tracking areas move with the satellites, a stationary UE would have to keep performing tracking area updates in idle mode, leading to a significant signaling load. One solution, which is being standardized in 3GPP, is to define tracking areas to be fixed on the ground to decouple from the radio cells.

In an LEO satellite access network, UEs in connected mode may need frequent handover. Conditional handover can be utilized to reduce handover signaling and increase handover robustness. In conditional handover, the network sends a handover command to the UE with a condition. The UE stores the command and applies the command when the condition is satisfied. Then the UE executes the handover and connects to the target cell. More tailored handover conditions can be introduced in the context of an LEO satellite access network, as specified in 3GPP. Example conditions may leverage the information about the UE location, satellite ephemeris, and the service duration of each LEO satellite over a particular area.

## IV. STANDARDIZATION ASPECTS

Most currently operating satellite networks are using a radio access technology based on European Telecommunications Standards Institute (ETSI) defined standards (e.g., Digital Video Broadcasting (DVB) or GEO-Mobile Radio Interface (GMR)) along with numerous proprietary radio protocols. None of these standards were sufficient to ensure interoperability between solution vendors and they did not lead to a sustainable ecosystem [15]. In contrast, terrestrial mobile networks have significantly benefited from international standardization efforts spearheaded by 3GPP over the past two decades. The outcome is a prosperous global mobile ecosystem with well-performing systems, products, and services.

Thanks to its inherent design flexibilities, 5G creates a unique opportunity to support satellite communications. With satellite access in the 3GPP roadmap, a global standard for satellite communications is being defined. It opens the door to integrating satellite components in 5G eco-systems, which will contribute to the delivery of 5G services, especially in areas beyond cellular coverage. Compared to the legacy value chain of satellite communications, this 3GPP-based standard will foster the development of smartphones and IoT devices with mobility and multi-connectivity capabilities between satellite and cellular access, which can enable:

- A seamless combination of satellite and cellular access for end-users
- Deployment and operation of multi-vendor satellite communication systems in the space segment while breaking the link between network infrastructure and terminal vendors
- Cost reduction through integration in the 3GPP eco-system addressing a global market and fostering economies of scale
- Native support of all 5G features (e.g., slicing, energy-saving, mobility, third-party network management, application & service platforms)

There have been several 3GPP initiatives on non-terrestrial networks (NTN), with LEO satellite access being a major component. The initiatives span across multiple areas from radio access network (RAN) to services and system aspects (SA) to core and terminals (CT). Figure 4 provides an overview of the satellite-related activities in the 3GPP roadmap up to the latest Release 17.

The work in 3GPP SA addresses both satellite access with a 3GPP-based air interface and satellite backhaul whose air interface can be 3GPP-based or non-3GPP-based. 3GPP RAN is adapting narrowband IoT (NB-IoT) and Long-Term Evolution (LTE) for machine-type communication (LTE-M) to support satellite access. This line of work is known as IoT NTN, which targets low-cost IoT devices with low service rate requirements. 3GPP RAN is also evolving the 5G New Radio (NR) air interface for satellite access in sub-6 GHz and higher frequencies. The operation in sub-6GHz aims to provide outdoor connectivity directly to handhelds, while the operation in higher frequencies aims to provide connectivity to VSAT terminals or ESIM terminals.

On the path to 6G, we anticipate that standardization will continue to be crucial to the success of satellite communication systems, including LEO satellite access. 3GPP has been continuously working on 5G evolution and will work on 6G in a few years. It is expected that the satellite industry will continue the standardization effort in 3GPP to advance the integration of satellites in 5G evolution and 6G systems over the subsequent releases of 3GPP. The continuous efforts will help ensure the introduction of new features to improve performance and provide new capabilities. Figure 4 provides our anticipation of the satellite-related standardization activities beyond the current 3GPP Release 17, including 5G Advanced and 6G.

## V. SATELLITE SERVICES AND BUSINESS ASPECTS

We will see a staggering increase in high-throughput satellite capacity in the coming years. By using the Ka-, Ku- and V-band spectrum and gigahertz wide bandwidth, speeds comparable to terrestrial LTE services may be offered by LEO satellite technology to connect devices such as VSAT and ESIM for fixed broadband connectivity services. However, such LEO-based fixed broadband connectivity services require devices equipped with custom-made flat panel or dish antenna featuring high gain and beam tracking capabilities. Solving these devices' cost and performance challenges is likely an equally big technical challenge as orbiting the satellite constellations. In areas where fiber or high-speed cable is available, they cannot compete. But outside those areas, LEO networks offering fixed broadband connectivity are likely competitive, also considering a faster time-to-service.

3GPP-based 5G and future 6G technologies can promise an open environment where devices can connect to 3GPP-based mobile broadband satellite systems. We expect that within several years, a large number of 5G devices with NTN support will be available, mainly achieved by software updates in chipsets and support for the necessary spectrum bands. On the path to 6G, end-user devices will have the unique capability to roam between cellular and satellite networks seamlessly, depending on agreements between local mobile network operators (MNO) and satellite network operators (SNO). Moreover, the use of 3GPP-based technology for LEO mobile broadband systems will facilitate the integration of satellite and terrestrial network technologies to improve coverage and increase resiliency.

While 3GPP-based LEO fixed broadband satellite access networks can be an alternative for offering fixed broadband connectivity to local area networks in areas where fiber or high-speed cable is not available, 3GPP-based LEO mobile broadband satellite access networks providing direct connectivity to handset or IoT devices will be a complement to provide the so often quoted 'ubiquitous' mobile service coverage. While terrestrial networks offer data rates of 100's of Mbps or even Gbps, it is expected that an LEO-based network providing direct connectivity can deliver speeds up to several Mbps to smartphones using 15 MHz of S-band spectrum and thus support voice, messaging, and basic data connectivity. Special devices with higher antenna gain or higher output power (e.g., ESIM and VSAT) can enjoy higher speeds. While a competitive LEO constellation providing direct connectivity to end-users for voice and data services will require a higher number of satellites, narrowband IoT type of services can be supported with a smaller investment for verticals such as transportation, logistics, utilities, farming, and mining.

The market for LEO-based direct connectivity networks can be divided into two segments. The first segment is guaranteed service with anywhere connectivity, where subscribers in developed markets are willing to pay a few percent extra on their monthly subscriptions. An allowed monthly 'NTN data limit' could be in the order of a few 100 MB per month. The other segment is uncovered areas in developing markets where subscribers will connect for basic communications. The average revenue per user (ARPU) in the latter segment is relatively small but still contributes to the overall revenue as LEO satellites constantly orbit around the Earth. Enterprise IoT connectivity stands for another promising opportunity where, e.g., containers, vehicles, and vessels could be connected. For LEO-based fixed broadband connectivity, prime business segments include maritime and aeronautical services, among others.

## VI. CONCLUSIONS AND 6G OUTLOOK

There have been enormous investments in LEO satellite access to cover the hardest to reach portions of our world to bridge the digital divide. As we have learned from history, realizing the success of space-based Internet services has not been – and will not be – easy. In this article, we have discussed some of the critical challenges, which are by no means exhaustive. We refer the interested readers to [2] for more discussions on the challenges.

6G research is ramping up. Extending the conventional terrestrial access to include LEO satellite access will be necessary to achieve truly global coverage. We anticipate that the integration of LEO satellite access with terrestrial networks will be tight in 6G, providing seamless mobility between terrestrial and LEO satellite networks. To achieve the tight integration in 6G, we will need to take a clean-slate approach by considering LEO satellite access in the design of the 6G system from day one, in contrast to the evolutionary approach undertaken in 5G NTN. The clean-slate approach opens the door to many research questions. It is expected that 6G standardization will start in 3GPP around 2025 and 6G commercialization will occur around 2030. We hold the view that there is some urgency to get going and the time to address the research questions towards 6G is now. We conclude by pointing out some fruitful avenues for 6G research.

*Waveform design*: 5G NTN is based on OFDM waveform to maximize synergies with existing terrestrial 5G systems. Considering the high time-varying Doppler shift in LEO satellite access and limited power resource of the satellite payload, OFDM may not be the optimal design choice (due to, e.g., peak-to-average-power-ratio issue). It will be interesting to study new waveforms that are more robust to time and frequency errors while being power efficient for the satellite payload.

*System architecture*: Designing a flexible network architecture for the integrated terrestrial and LEO satellite access will be imperative. The architecture should support efficient load balancing, seamless mobility, and dynamic resource allocation between the terrestrial and LEO satellite access. In addition, the architecture should be flexible enough to support ease of constellation management and the diverse LEO satellite deployment scenarios with ISLs, multi-link connectivity, backhauling, etc.

*Spectrum usage:* The current spectrum allocations for terrestrial and satellite communications are relatively static. Interference coordination of the two types of systems is mainly governed by regulatory requirements. With tight integration of terrestrial networks with LEO satellite access in 6G, spectrum usage may become more flexible, leading to more optimized interference coordination and improved spectrum efficiency.

*Trials and test deployments:* We have a window of opportunity between now and the start of 6G standardization around 2025 to conduct early trials and test deployments of integrated terrestrial and LEO satellite access. The trials and test deployments will provide valuable practical knowledge, which will guide the design of the 6G system featuring tight integration of terrestrial networks with LEO satellite access.

## BIOGRAPHIES

**Xingqin Lin** is a Master Researcher at Ericsson, driving non-terrestrial networks research and standardization.

**Stefano Cioni** is a Telecommunication Systems Engineer at the European Space Agency.

**Gilles Charbit** is a Distinguished Member of Technical Staff at MediaTek and the rapporteur for the 3GPP Release-17 IoT NTN work item.

**Nicolas Chuberre** is a Solution Line Manager at Thales Alenia Space and the rapporteur for the 3GPP Release-17 NR NTN work item.

**Sven Hellsten** is a Business Development Director at Ericsson.

**Jean-Francois Boutillon** is a Broadband Constellation Solution Line Manager at Thales Alenia Space.